\documentclass{ws-mpla}
\usepackage[super]{cite}
\usepackage{graphicx}
\usepackage{amsmath}
\usepackage{amssymb}
\usepackage{graphicx}
\usepackage{color}
\usepackage{nicefrac}
\usepackage{url,overcite}
\usepackage{subfigure}

\newcommand{\note}[1]{\emph{\textcolor{red}{}}}

\newcommand{\Msun}{{\ensuremath{{M}_{\odot}}}}

\newcommand{\K}{\ensuremath{\mathrm{K}}}

\newcommand{\gcc}{\ensuremath{\mathrm{g}\,\mathrm{cm}^{-3}}}

\newcommand{\Ni}{{\ensuremath{^{56}\mathrm{Ni}}}}

\newcommand{\Ox}{{\ensuremath{^{16}\mathrm{O}}}}

\newcommand{\Si}{{\ensuremath{^{28}\mathrm{Si}}}}

\newcommand{\Cx}{{\ensuremath{^{12}\mathrm{C}}}}

\newcommand{\MS}{{\ensuremath{M_*}}}
\newcommand{\MHe}{{\ensuremath{M_{\mathrm{He}}}}}
\newcommand{\rhoc}{{\ensuremath{\rho_{\mathrm{c}}}}}
\newcommand{\Tc}{{\ensuremath{T_{\mathrm{c}}}}}

\def\gb {{\bf g}}

\def\ub {{\bf u}}

\newcommand{\Cplusplus}{{\rmfamily C\raise.22ex\hbox{\small ++} }}

\newcommand{\lSect}[1]{\label{#1}}

\newcommand{\Sect}[1]{{\S~\ref{#1}}}

\newcommand{\Ep}[1]{{\ensuremath{10^{#1}}}}

\newcommand{\cm}{{\ensuremath{\mathrm{cm}}}}

\newcommand{\erg}{{\ensuremath{\mathrm{erg}}}}

\newcommand{\CASTRO}{\texttt{CASTRO}}
\newcommand{\KEPLER}{\texttt{KEPLER}}

\newcommand{\araa}{ARA\&A}%
\newcommand{\apj}{ApJ}%
\newcommand{\apjl}{{ApJ}}%
\newcommand{\apjs}{{ApJS}}%
\newcommand{\aap}{{A\&A}}%
\newcommand{\mnras}{{MNRAS}}%
\newcommand{\nat}{{Nature}}%
\newcommand{\sfrac}[2]{\mathchoice
  {\kern0em\raise.5ex\hbox{\the\scriptfont0 #1}\kern-.15em/
   \kern-.15em\lower.25ex\hbox{\the\scriptfont0 #2}}
  {\kern0em\raise.5ex\hbox{\the\scriptfont0 #1}\kern-.15em/
   \kern-.15em\lower.25ex\hbox{\the\scriptfont0 #2}}
  {\kern0em\raise.5ex\hbox{\the\scriptscriptfont0 #1}\kern-.2em/
   \kern-.15em\lower.25ex\hbox{\the\scriptscriptfont0 #2}}
  {#1\!/#2}}

\def\gb {{\bf g}}

\def\ub {{\bf u}}

\newcommand{\be}{\begin{equation}}
\newcommand{\ee}{\end{equation}}
\newcommand{\bea}{\begin{eqnarray}}
\newcommand{\eea}{\end{eqnarray}}
\newcommand{\bc}{\begin{center}}
\newcommand{\ec}{\end{center}}

\begin{document}

\markboth{Chen}
{Pair-Instability Supernovae of Fast Rotating Stars}

\catchline{}{}{}{}{}

\title{Pair-Instability Supernovae of Fast Rotating Stars}

\author{Ke-Jung Chen}
\address{Department of Astronomy \& Astrophysics, University of California, 1156 High St.\\ 
Santa Cruz, California 95064, USA \\ 
School of Physics and Astronomy, University of Minnesota, 116 Church St. \\
Minneapolis, Minnesota 55455, USA \\ 
kchen@ucolick.org
}

\maketitle


\begin{abstract}
We present 2D simulations of pair-instability supernovae considering rapid rotation during 
their explosion phases. Recent studies of the Pop~III star formation suggested that these stars could be 
born with a mass scale about 100 \Msun\ and with a strong rotation.  Based on stellar evolution models,
these massive Pop~III stars might have died as highly energetic pair-instability supernovae. 
We perform 2D calculations to investigate the impact of rotation on pair-instability supernovae. 
Our results suggest that rotation leads to an aspherical explosion due to an anisotropic collapse. 
If the first stars have a $50\%$ of keplerian rotational rate of the oxygen core before their pair-instability
 explosions, the overall \Ni{} production can be significantly reduced by about two orders of magnitude. 
 An extreme case of $100\%$ keplerian rotational rate shows an interesting feature of fluid
 instabilities along the equatorial plane caused by non-synchronized and non-isotropic 
 ignitions of explosions, so that the shocks run into the in-falling gas and generate the 
 Richtmyer--Meshkov instability.  


\end{abstract}

\ccode{PACS Nos.: include PACS Nos.}
\section{Introduction}
\lSect{introduction}
The first generation of stars so-called, Population~III (Pop~III) stars were predicted to form about several 
hundred million years after the Big Bang [\refcite{abel2002,bromm2009}]. The Pop~III stars were born 
in the gravitational-potential wells constructed by the dark matter halos of masses about $10^6$ solar masses 
(\Msun), that allowed the primordial gas (mass content: 76 \% hydrogen, 24\% helium) to form stars. 
Because no heavy elements (metals) were present at that time, molecular hydrogen served as the most 
efficient coolant, but it could not cool the gas temperature  efficiently. Thus, the Jeans mass of the 
Pop~III star-forming cloud was expected to be  much larger than the one of the current star formation regions; 
consequently the Pop~III stars could be more massive than the present-day stars. Modern cosmology
 simulations suggested that the mass scale of the Pop~III stars would fall between tens 
to hundreds of \Msun. The exact mass scale is still under debate. 
Without initial metals such as carbon and oxygen, the Pop~III stars needed to reach higher 
core temperatures to burn hydrogen effectively [\refcite{schaerer2002}], that also resulted in higher 
surface temperatures.  So these stars could produce large amounts of UV photons and extensively 
ionized primordial hydrogen, and helium inside the  inter-galactic medium (IGM) leading to the cosmic
 reionization [\refcite{carr1984,whalen2004,kitayama2004,alvarez2007,abel2007}].  
After their short life time of several million years, many massive Pop~III stars might have died as supernovae 
and dispersed the first metal to the primordial IGM. Such chemical enrichment might have changed the later star 
formation mode from Pop~III to Pop~II and profoundly reshaped the simple early universe into a state of 
ever-increasing complexity. Thus, understanding the first stars and their supernovae has become one of the 
frontiers of modern cosmology.  

Fate of a massive star strongly correlate to its final mass before its death. Because there is no initial metal 
inside the envelope of Pop~III stars, the mass loss rates due to the metal line-driven wind can be 
strongly suppressed. Hence, it is reasonable to assume that Pop~III stars losses very little or no mass 
during their stellar evolution. Based on stellar physics [\refcite{kipp1990,woosley2002}], the Pop~III 
stars with initial masses of $10 - 80\,\Msun$ eventually forge an iron core about one solar mass. When 
the iron core mass exceeds its Chandrasekhar mass [\refcite{chand1942}], the degenerate pressure of 
electrons is no longer able to support its own gravity. The iron core catastrophically collapses into a black hole 
or neutron star. The energy released from the gravity allows an energetic explosion so-called a core-collapse 
supernova. However, such explosions may fail and the entire star may collapses  directly into a black hole.
The nature of core-collapse supernovae is complicated by several hurdles such as neutrino physics, 
multi-scale, and multi-dimension [\refcite{burrows1995,janka1996,mezz1998,murphy2008,nordhaus2010}].  
The explosion mechanisms  remain subject to investigation.  If Pop~III stars are more massive than $80 \Msun$. after 
the central carbon burning, their cores encounter the electron/positron production instability,  the core 
reaches sufficiently high  temperatures ($\sim \Ep9\,\K$) and low densities ($\sim\Ep6\,\gcc$) to  
favor the creation of electron-positron pairs. It reduces the adiabatic index $\gamma$ below 4/3 and  
causes a dynamical instability of the core. The core is now at quasi hydrodynamical equilibrium and its
temperatures begin to oscillate for a period of its dynamic time scale of several hundred seconds. The 
variations in temperature do not cause a significant impact to its evolution and the star 
later dies as a core-collapse supernova. However, if the stars are more than 100 \Msun, the 
oscillation of temperatures becomes very violent.  Several large temperature spikes produce shocks 
with energy of $10^{49} - 10^{50} \erg$. These shocks can not disperse the entire star, but eject its
outer envelopes instead.  Due to different amount of mass and energetics, these 
ejected shells are likely to run into each other. The colliding shells effectively convert the kinetic 
energy of ejecta into the internal energy of gas. The collision sites usually happen at optically thin regions
and thermal radiation escape immediately and produce very luminous optical transits known as pulsational
pair-instability supernovae [\refcite{woosley2007,chen2014b}]. Once the mass of stars reaches between 150  
and 260 \Msun. At this point, pair-instabilities become so violent and trigger a runaway collapse of core. The 
core temperature and density rise swiftly and ignite the explosive oxygen/silicon burning. The energy released 
from the burning converts the contraction of the core into an energetic explosion that completely disrupts the entire 
star.  This thermonuclear explosion, known as a pair-instability supernova (PSN), that has explosion energies 
up to $\Ep{53}\,\erg$ and yield up to 50 $\Msun$ of \Ni{} 
[\refcite{barkat1967,glatzel1985,heger2002,heger2010,kasen2011,chen2011,chen2014c}].  
Most of Stars are  300 \Msun are believed to simply die as a black hole. A new type of supernova caused 
by the general-relativity instability in supermassive stars at $ \sim 55,000$ \Msun, which may have 
formed in the early universe, has now also been found [\refcite{chen2014d}]. 

Because the Pop~III stars were predicted to very massive, many of them tended to die as PSNe.
The 1D PSN models have been extensively studied by several authors [\refcite{heger2002,heger2010,kasen2011,dessart2013}]. 
The extensive yields, explosion energies, and radiation properties of PSNe are calculated in their models. 
Unlike core-collapse 
supernovae, the thermonuclear explosions of PSNe are not very sensitive to the dimensionality of simulations. However, 
simulating the mixing of the fluid instabilities in PSNe still requires multidimensional simulations that recently become accessible and 
studied by [\refcite{candace2011,chen2011,chatz2012a,chen2014b}]. The simulations of [\refcite{chen2014a}] found the mixing of 
PSN from the both burning and hydrodynamics instabilities. In the red super-giant progenitors, the mixing driven by the 
Rayleigh-Taylor instabilities can be very significant.

One of the important physical quantities in the stellar evolution is rotation. In the real world, the process of star 
formation usually endows a certain amount of angular momentum to the star and makes it rotate. Due to the technical 
difficulty of the study, the rotational rates of massive stars are still poorly understood in both theoretical and observational 
studies. Rotation can affect the stellar evolution as well as the resulting supernova explosions. The nucleosynthesis 
inside the stars changes due to the rotational mixing, so that the chemical composition of the star changes. 
If metal from the inner part is mixed out to the envelope,  the stellar wind can be enhanced due to metal-driven lines. 
Strong winds reduce the stellar mass dramatically and affect the fate of the stars. 
Studies of the stellar evolution of very massive stars with rotation have been performed 
by [\refcite{glatzel1985,maeder2000,heger2005,hirschi2007,eks2008}].  Recent results 
from [\refcite{chatz2012b,yoon2012}] show that rotation can lower the mass criterion for PSNe 
progenitors because mixing facilitates helium burning, resulting in a more massive oxygen core. 
Recent simulations [\refcite{stacy2011}] suggested that the Pop~III stars could rotate very fast, up to $50\,\%$ of 
keplerian rate at the surface. If such a high rotation does exist, it can affect the evolution of the stars and their 
supernovae. So we perform 2D simulations of PSNe with rotation and investigate how the 
strong rotation impacts the explosive burning. The purpose of this paper is to explore the energetics and nucleosynthesis 
of PSNe during the explosion phase. 

The structure of this paper is as follows: we first describe the numerical approaches and setup of our simulations 
in \Sect{rot_methodology}. Then we present the results in \Sect{rot_results} and conclude our findings in \Sect{rot_discussion}. 

\section{Methodology and Problem Setup}
\lSect{rot_methodology}
The 3D stellar evolution models followed from the main sequence star 
to supernovae are still unavailable due to the limitations of current computational power. Alternatively, 
we start simulating the stars with \KEPLER{} [\refcite{kepler,heger2001}], a 1D spherically-symmetric 
Lagrangian code, and follow the evolution of stars up to $20-100$ seconds before the supernova explosions. 
Then we map the resulting 1D stellar evolution models from \KEPLER{} onto 2D grids of \CASTRO{} as the initial 
conditions. We then apply our differential rotation scheme to \CASTRO{} and follow the simulations until the 
explosive burning ceases when the shock has successfully launched. This setup is designed to model the most 
critical phases of the supernova explosion in 2D with practical computational resources. In this section, we introduce 
our progenitor models, problem setup, and numerical methods.  

\subsection{Progenitor Models}
We use progenitor Pop~III stars of $150\,\Msun$, $200\,\Msun$, and $250\,\Msun$ with very
little overshooting at the late-time evolution. These stars eventually become blue supergiants 
which have smaller radii than those of red supergiants. The 1D stellar models are evolved until 
most of the explosive burning is about to happen. Physical properties of these stars are listed 
in Table 1. The resulting 1D profiles are then mapped onto 2D cylindrical grids of \CASTRO{} 
using a new mapping algorithm [\refcite{chen2013}]. This initialization scheme conservatively maps  
the physical quantities such as mass and energy during mapping from 1D to multi-D. The initial 
perturbations are seeded in the form of velocities based on the stellar convection physics.

\begin{table}[h]
\tbl{1D Progenitor models}
{\begin{tabular}{@{}lccccc@{}} \toprule
Name &  $\MS$ & $\MHe$ & $\rhoc$ &  $\Tc$ & $R$    \\  
{}    & {[$\Msun$]} &  {[$\Msun$]} &  {[$\Ep6\gcc$]} &  {[$\Ep9\,\K$]} & {[$\Ep{13}\,\cm$] } \\
\colrule
B150  & 150 & 67  & 1.40 & 3.25 & 16.54   \\
B200  & 200 & 95  & 1.23 & 3.31 &  2.86   \\
B250  & 250 & 109 & 1.11 & 3.34 & 23.06  \\
\botrule
\end{tabular}\label{ta1} }
\begin{center}
{\small $\MS$: initial stellar mass; $\MHe$: helium core mass at collapse; $\rhoc$: central density at collapse; 
$\Tc$: central temperature at collapse; and $R$: stellar radius.}
\end{center}
\end{table}

\subsection{\CASTRO}
We run 2D simulations using \CASTRO{} [\refcite{ann2010,zhang2011}],  
a massively parallel, multidimensional Eulerian, adaptive mesh refinement (AMR),
hydrodynamics code for astrophysical applications. \CASTRO{} was originally developed at 
the Lawrence Berkeley Lab, and it is designed to run effectively on supercomputers over 
10,000 CPUs. \CASTRO{} provides a powerful platform for simulating hydrodynamics and gravity
for astrophysical gas dynamics. Modeling thermonuclear supernovae requires calculating the 
energy generation rates from nuclear burning, which occurs over an extensive range of temperatures, 
densities, and compositions of isotopes. We have implemented the APPROX 19-isotope reaction network
[\refcite{kepler,timmes1999}] into \CASTRO.   It is capable of efficiently calculating accurate energy 
generation rates for  major nuclear reactions from hydrogen to silicon burning.  In \CASTRO, we 
use a realistic equation of state (EOS) [\refcite{timmes2000}] for stellar matter. This EOS considers 
the (non)degenerate and (non)relativistic electrons, electron-positron pair production, 
as well as ideal gas with radiation. It is a tabular EOS that reads in $\rho$, $T$, and $X_i$ of gas and 
yields its derivative thermodynamics quantities. \CASTRO{} offers different types of calculation for gravity, 
including Constant, Poisson, and Monopole.  For supernova simulations, spherical symmetry is still a good 
approximation for the mass distribution of gas at the early phase of evolution.  Using this approximation 
saves lot of computational time in the gravity solver. For 2D or 3D \CASTRO{} simulations, we first calculate a 
1D radial average profile of density, then compute the 1D profile of $\gb$ field and use it to calculate the gravity 
of the multidimensional grid cells.  Care is taken to resolve the important scales of the explosions such as 
catching the shock front and mixing driven by fluid instabilities. The grid structure for 2D simulations uses 
base grids, $256\times512$ with additional three levels of AMR to resolve a domain of 
$(4\times10^{11})\times(8\times10^{11})\,\cm^2$. This setup yields the size of the finest patch about 
$(7.2\times 10^7)^2\,\cm^2$ which fully resolves the nuclear burning and the structures of fluid instabilities of 
rotating PSNe. The AMR criteria are set for gradients of density, velocity, and pressure. Finer zones are 
automatically created on the top of over-gradient regions. 2D \CASTRO{} uses a cylindrical 
coordinate $r-z$. The axis $z$ serves as the rotational axis. Since we simulate only half of the star in 2D. 
The lower boundary of $r$ uses reflect conditions to prohibit fluid from entering; the other three boundaries 
use outflow conditions that allow the fluid to freely cross over. We also constructed nested zones to ensure 
that the inner region of the star constantly receives a higher spatial resolution.

\subsection{2D Rotational Model}
\begin{figure}[h]
\begin{center}
\includegraphics[width=.6\columnwidth]{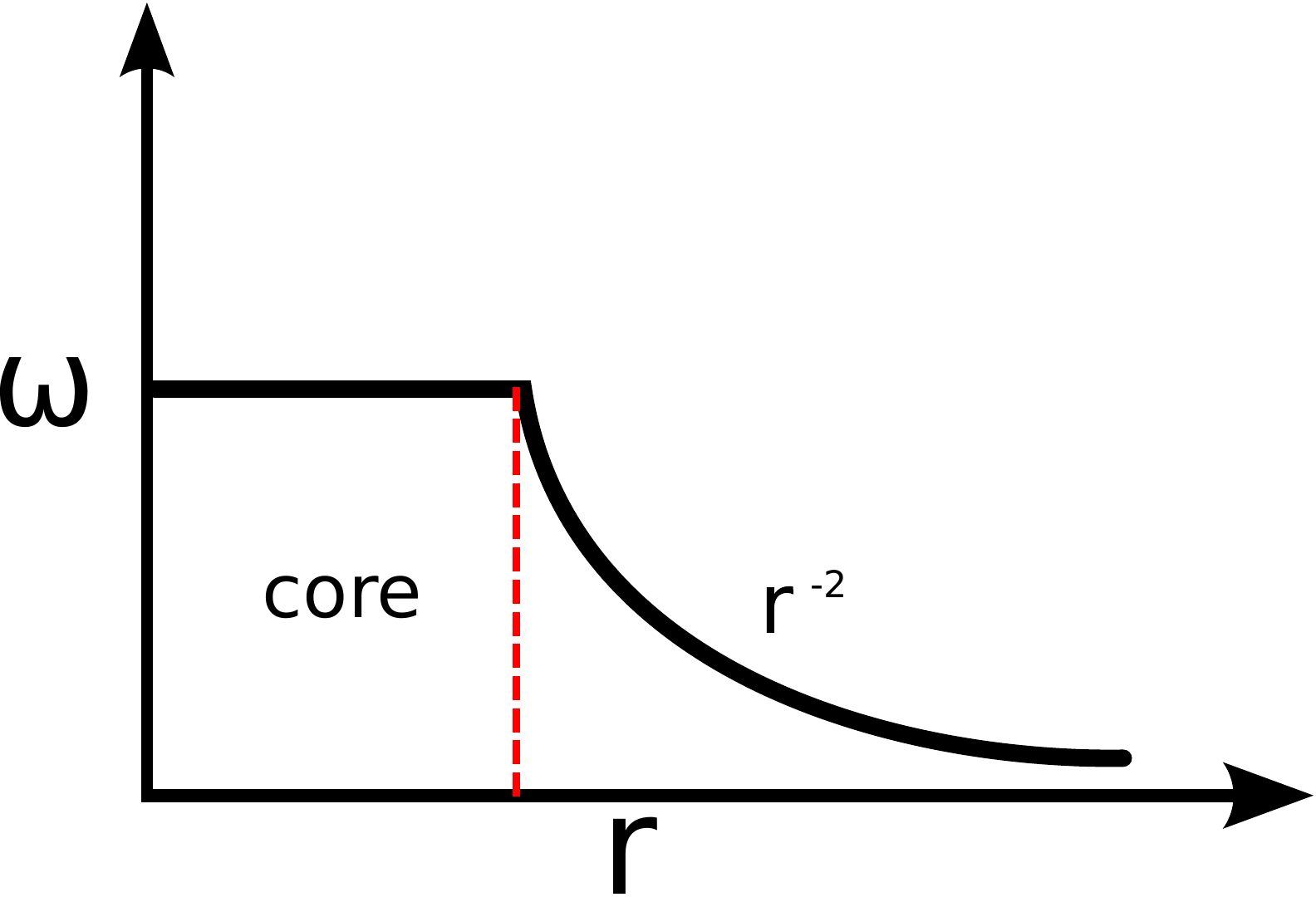} 
\end{center}
\caption[]{Illustration of the differential rotational model. $\omega$ is the rotational rate and $r$ is the distance to 
the rotational axis. $\omega$ inside the oxygen core ($r\,\leq r_c $) is assumed as a constant. Beyond the core  
($r\,>\, r_c $), we assume a specific angular $j$ constant and result in $\omega \propto r^{-2}$. \label{rot_model}}
\end{figure}

\begin{figure}[h]
\begin{center} 
\subfigure[Oxygen abundance pattern]{\label{osize}\includegraphics[width=.48\columnwidth]{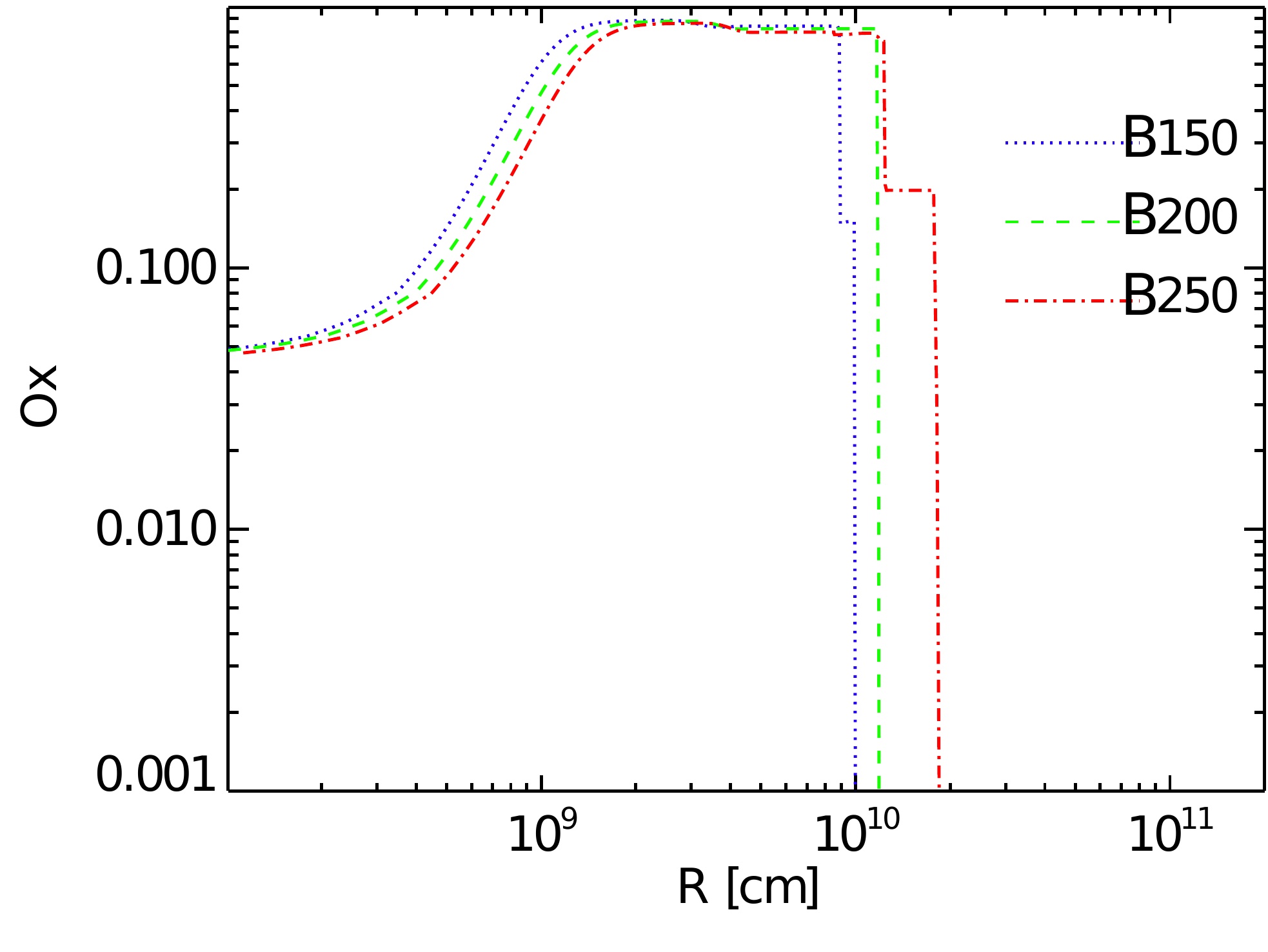}}                
\subfigure[ $\omega_k$ as a function radius]{\label{somega} \includegraphics[width=0.48\textwidth]{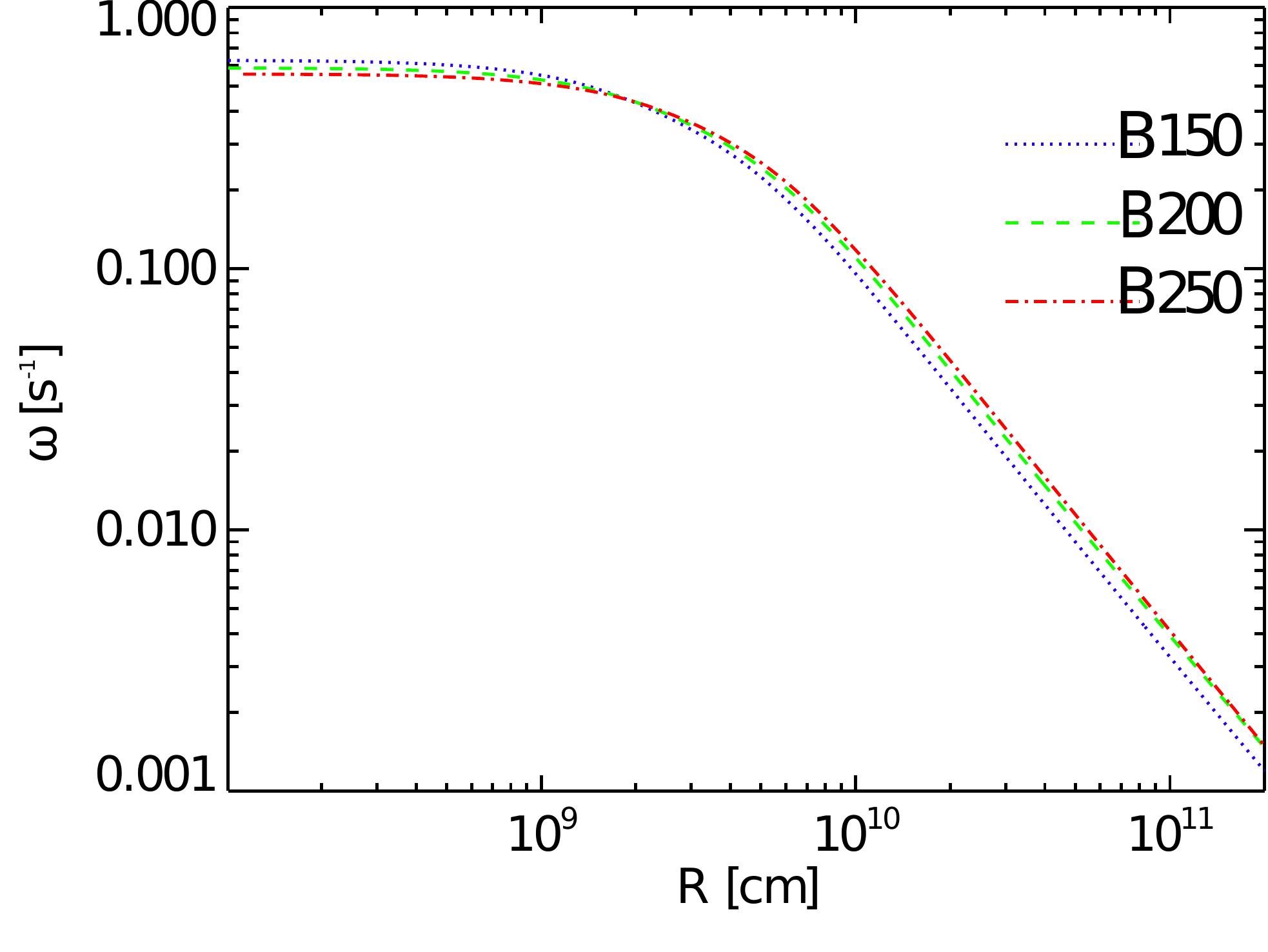}}   
\caption[The rotating oxygen]{(a) The size of the oxygen cores of presupernovae can be identified by their oxygen abundance. 
The sudden drop in the oxygen core marks the size of oxygen core. (b) $\omega(r)$ behaves like a 
constant then decreases.  The saddle point of the drop is roughly at the edge of oxygen core. \label{rot_oxygen}}
\end{center}
\end{figure}

We develop a differential rotational model motivated by the stellar structure. It yields non-uniform 
rotational rates inside different regions of stars. Because the oxygen core of the star is a dense and compact object, we assume 
it behaves like a rigid body with a constant rotational rate, $\omega$.  Outside the oxygen core, the structure 
of the helium and hydrogen envelope is relatively puffy and unlikely to keep the constant $\omega$. Instead, 
we assume a constant specific angular moment extending from the edge of oxygen core to its surface and 
$\omega$ monotonically decreases from the rotational axis, as shown in Figure~\ref{rot_model}. To determine 
the radius of oxygen core and its rotational rate, we plot the oxygen abundance as a function of radius in 
Figure~\ref{osize}. The most abundant oxygen is located at the radius between $10^9-10^{10}\,\cm$, and its outer 
boundaries of the oxygen core is at $10^{10}\,\cm$. Based on the size and mass of the oxygen core, we 
determine its keplerian rotational rates, $\omega_k$ which is about $0.5\,\sec^{-1}$.  To examine whether the 
model is viable, we calculate $\omega_k$ as a function of radius in Figure~\ref{somega}. It shows a nice trend of constant  
$\omega$ in the the oxygen core and slowly decrease from the core boundary. Our assumption of oxygen rotating as a 
rigid body seems to be very suitable for the model. 

We therefore assume a constant rotational rate inside the oxygen core and this rate is based on the keplerian 
rate of the oxygen core. In \CASTRO, the rotation is calculated by the angular  momentum, $j$ which 
provides a force term in the Euler equation for fluids and itself is evolved by advect equations:
\begin{eqnarray}
\frac{\partial (\rho \ub)}{\partial t} &=& - \nabla \cdot (\rho \ub \ub) - \nabla p +\rho \gb + {\bf F}_c, \\
\frac{\partial (\rho j)}{\partial t} &=& - \nabla \cdot (\rho \ub j). \\
{\bf F}_c &=& \rho \frac{j^2}{r_{z}^{3}}; \quad \mbox{ $j = \omega r^2$ when $r \leq r_{c}$, otherwise $j = \omega r^2_c$  }
\end{eqnarray}
Here $\rho, p, \ub, \gb $ are the density, pressure, velocity vector, and gravitational vector, respectively. 
${\bf F}_c$ is the centrifugal force calculated by $\rho$, the angular momentum per unit mass  $j$,  and 
the distance to the rotational axis $r_z$. $j$ is only initialized at the beginning of the simulations, 
then it evolves with fluid elements following the equations as above. 
\section{Results}
\begin{figure}[h]
\begin{center}
\includegraphics[width=1.\columnwidth]{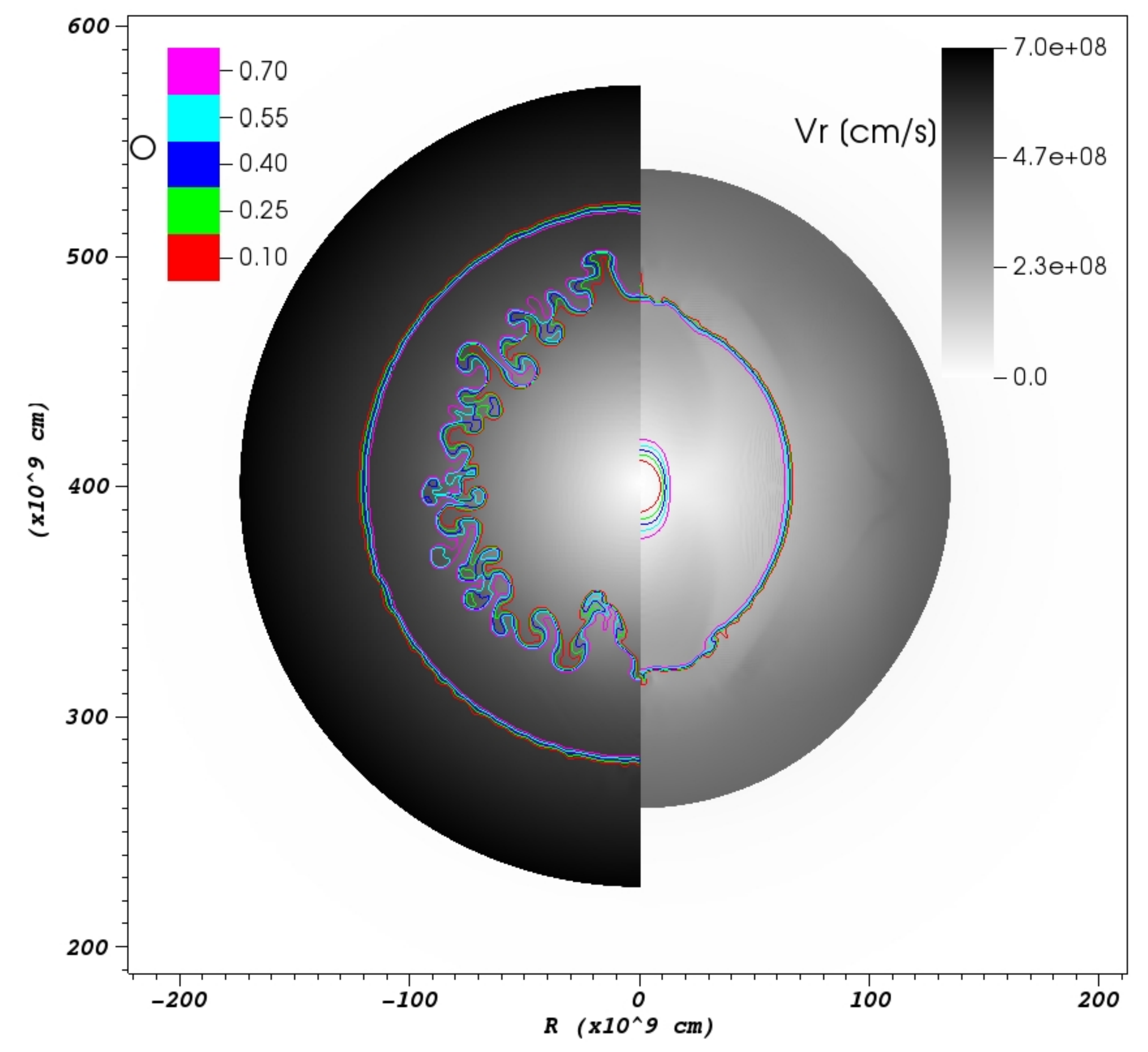} 
\end{center}
\caption[Comparison of rotating and non-rotating models I]{Comparison of rotating and non-rotating models from a $200\,\Msun$ Pop~III 
star. The left-half sphere is the non-rotating model, and the right-half is the rotating model of $\omega \sim 0.5 \omega_k$. The gray 
colors show the radial velocities $v_r$, and the colorful contours are the \Ox{} mass fraction. Both snapshots are taken at the same time, 
about $200\,\sec$ after the onset of the explosion. \label{rot_2d_a} }
\end{figure}

\begin{figure}[h]
\begin{center}
\includegraphics[width=1.\columnwidth]{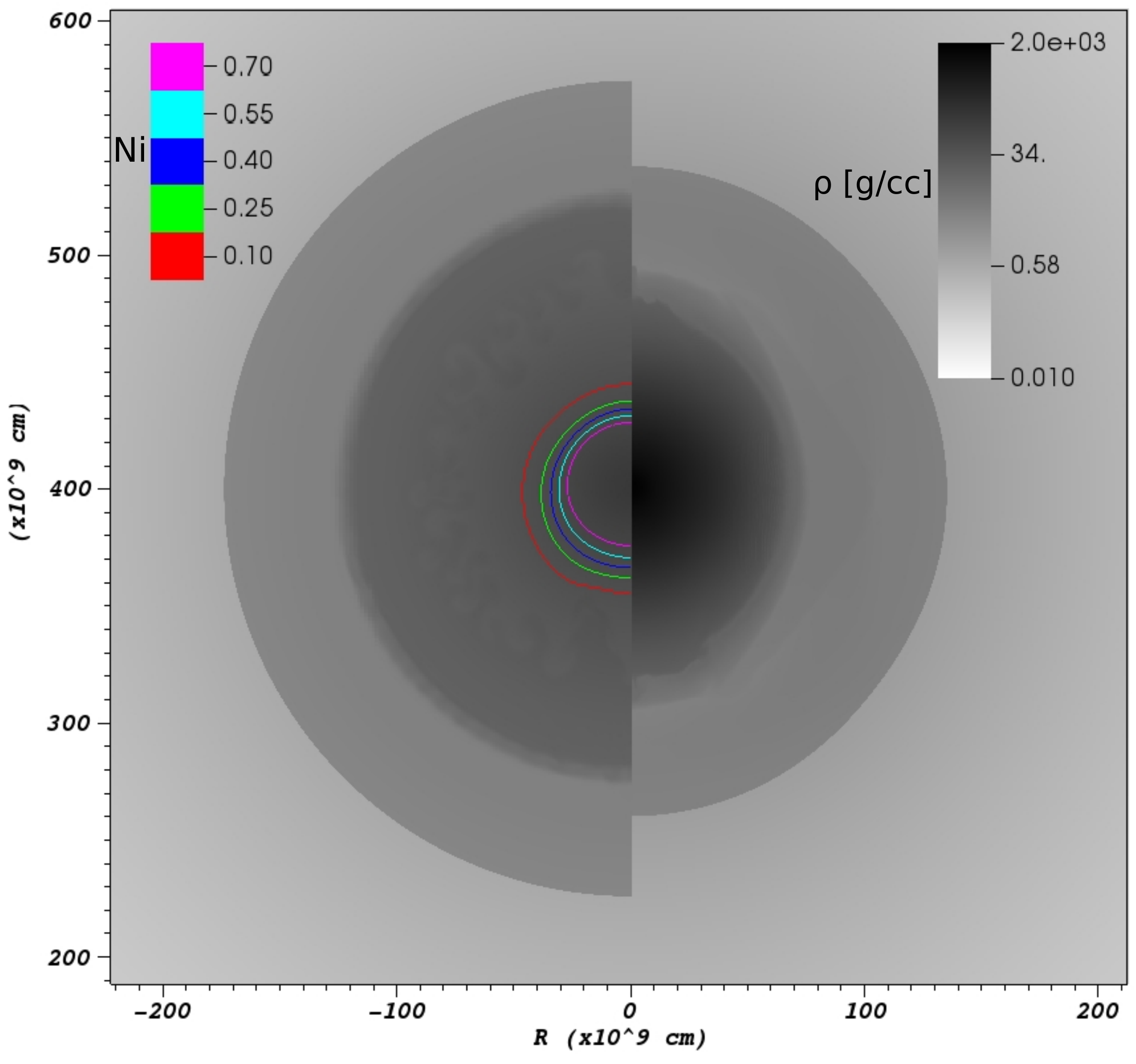} 
\end{center}
\caption[Comparison of rotating and non-rotating model II]{Comparison of rotating and non-rotating models from a $200\,\Msun$ 
star. The left-side sphere is the non-rotating model, and the right side shows a rotating model of $\omega \sim 0.5 \omega_k $. The gray 
color shows the density, and the colorful contours are the \Ni{} mass fraction. Both snapshots are taken at the same time, about $200\,\sec$ 
after the onset of the explosion. \label{rot_2d_b}}
\end{figure}

\lSect{rot_results}
We first present the results from the representative model of the $200\,\Msun$ rotating Pop~III star with $\omega = 0.5\omega_k$.
After the onset of the 2D \CASTRO{} simulation, the core of the star starts  
to collapse due to its dynamical instabilities caused by pair-production instabilities.  Similar to the case of
non-rotating stars, explosive oxygen and silicon burning proceed as the core contracts.  After about $25$ seconds, 
the energy released from the nuclear burning launches a shock and the core of star has successfully exploded.
For a non-rotating case, collapse of the core is perfectly anisotropic and homogeneous. Thus, the resulting 
explosion appears to have spherical symmetry.  However, for a rotating star,  rotation provides a centrifugal force to 
resist the in-falling  gas during the collapse.  Because the centrifugal force follows as ${\bf F}_c\propto r_{z}\omega^2$, 
for the given distance from the center of star,  the core receives the strongest ${\bf F}_c$ along the equatorial direction 
and no force in the polar direction. It automatically leads to anisotropic compression of the core and reflects in the 
explosion. The maximum compression happens along the polar direction and yields more explosion energy and 
releases a more powerful shock wave than other directions. The core is blew up in an elliptical shape with a long axis along the 
rotational axis. Figure~\ref{rot_2d_a} shows the post-explosion of  $200\,\Msun$ progenitor star, $\omega = 0$ in the left-half panel and $\omega = 0.5\omega_k$ in the 
right-half panel. Both snapshots are taken at the same time after the explosions occur. In comparison of both results, the non-rotating 
model shows a stronger shock wave, and demonstrates some mixing at the inner part of the oxygen-burning shells. The rotating model, 
however, shows a relatively less energetic explosion due to a weaker compression; the ejected oxygen-burning shell is shaped into  
an ellipse and shows very little mixing. PSNe of non-rotating stars usually produce a large amount of radioactive isotope, \Ni\ which 
is made inside PSNe mainly by the explosive \Si{} burning right before the core bounces. Because the decay energy from \Ni{} powers 
the PSN light curves,  it becomes one of the key isotopes for observations. The amount of \Ni{} production can determine whether the PSNe become
bright or faint SNe.  We show the corresponding \Ni{} abundance and gas density in Figure~\ref{rot_2d_b}. The non-rotating star
demonstrate a visual \Ni\ abundance, which is missing  in the rotating model.  It suggests that \Ni{}  production has been significantly 
suppressed due to rotation. Only about $10^{-2}\,\Msun$ \Ni{}  is made in the rotational model. However, the non-rotating model has 
synthesized about $6.57\,\Msun$ of \Ni. This suggests that a strong rotation can significantly affect the energetic and \Ni{} production in PSNe .

\begin{figure}[h]
\begin{center}
\includegraphics[width=1.\columnwidth]{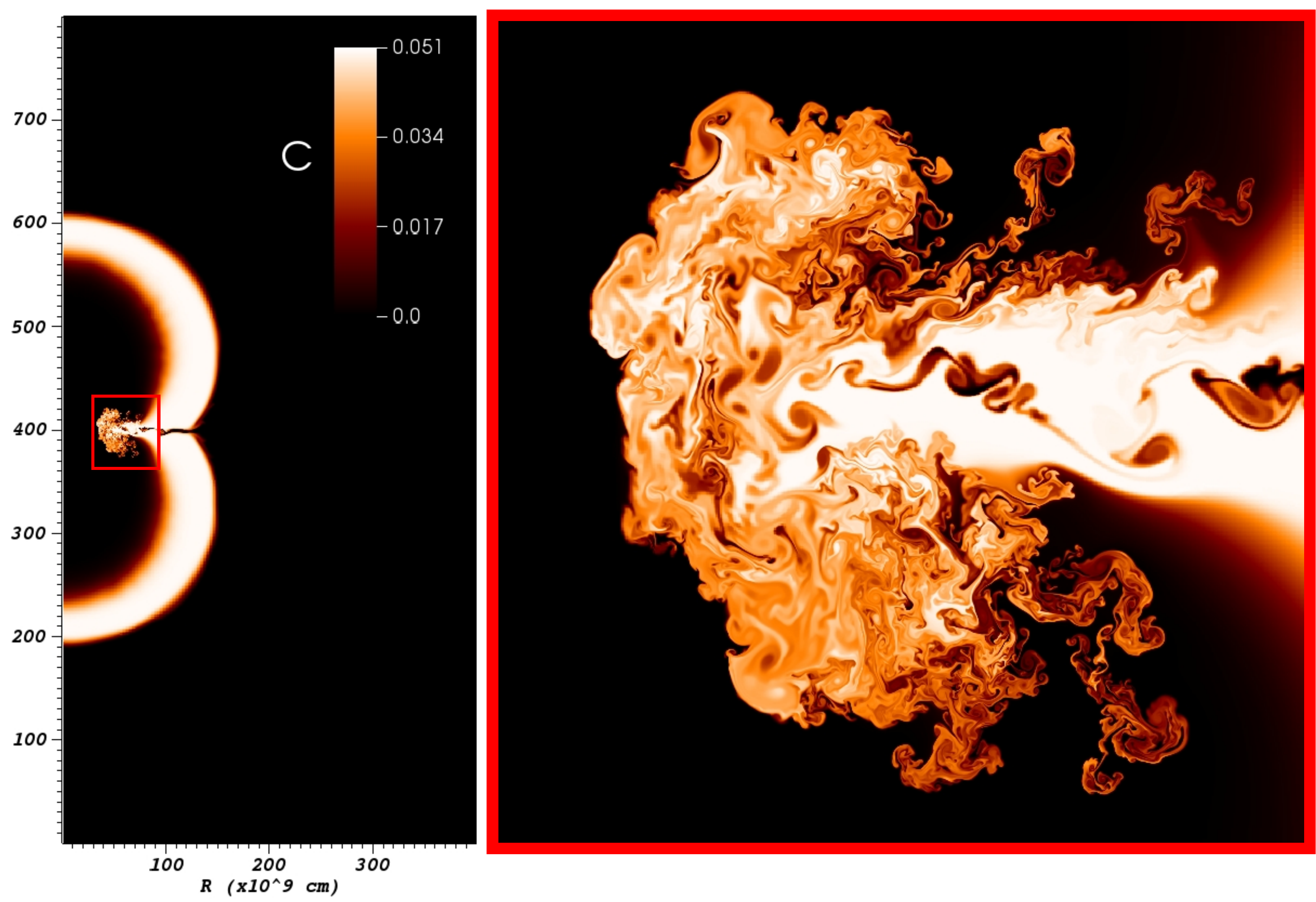} 
\end{center}
\caption[Overshooting]{Hot-color shows the mass fraction of $\Cx$. Strong fluid instabilities 
occur during the explosion along the equatorial plane shown in the red box of the left panel; 
its close-up appears in the right panel. 
\label{overshooting}}
\end{figure}

What happens if $\omega = \omega_k$ ? This is an extreme rotational rate which can almost break out of the star's core. To answer this question, we simulate 
a $200\,\Msun$ of $\omega = \omega_k$. The anisotropic compression becomes even more stronger than that of 
the $\omega = 0.5\omega_k$ case. 
The ellipse shape of the ejected core has an larger eccentricity. One interesting fluid instability has been found in the carbon-burning 
shell, which is shown in Figure~\ref{overshooting}. The carbon shell breaks along the equatorial plane and overshoots some carbon 
into the oxygen core. This overshooting develops strong fluid instabilities. Because the explosion is anisotropic, the shocks of different 
directions are initialized at different  times: the pole comes first, and the equator comes last. In this fast-rotating model, the 
shock from the pole has been sent out, but the gas along the equator is still collapsing. 
Once the shock runs into the collapsing gas, the Richtmyer$-$Meshkov (RM) instability [\refcite{rm}] starts to develop and the resulting 
mixing makes carbon penetrate into the oxygen-rich envelope.

\section{Conclusions}
\lSect{rot_discussion}
Stellar rotation plays a key role in both stellar evolution and the fate of the very massive Pop~III stars. 
The nature of the rotation can be traced back to the phase of star formation, when they are endowed with 
angular momentum. The definite rotational rates for the massive Pop~III stars are still unknown. Results of 
these star formation suggest that a high rotational rate is $50\%$ of its keplerian rotational rate. We present the results of  
the impact of rotation on a $200\,\Msun$ star.  The $50\%$ keplerian rotational rate case demonstrates the onset of an 
anisotropic explosion and strongly suppresses  \Ni{}. If this does apply for PSN progenitors,  \Ni{} production 
shown in previous PSN models may change and the corresponding PSNe luminosity  will be attenuated. 
An extreme case of a $100\%$ keplerian rate shows an interesting feature of overshooting along the equatorial 
plane caused by non-synchronized explosions that send shocks into the infilling gas. Finally, we conclude that 
PSNe of fast rotating stars would look much fainter than we originally expect because of the limited \Ni\ production.

\section*{Acknowledgments}
I thank Stan Woosley, Ann Almgren, Weiqun Zheng, Alexander Heger, Dan Whalen, and Volker Bromm 
for many useful discussions. Work at UCSC has been supported by an IAU-Gruber Fellowship, 
the DOE HEP Program under contract DE-SC0010676 and the NASA Theory Program  (NNX14AH34G). 
\CASTRO{} was developed through the DOE SciDAC program 
through grants; DE-AC02-05CH11231, and DE-FC02-09ER41618. All numerical simulations were 
performed with allocations from the National Energy Research Scientific Computing Center (NERSC) 
and the Minnesota Supercomputing Institute (MSI) .


\end{document}